\documentclass[prb,aps,twocolumn,amsmath,amssymb,floatfix,showpacs,
superscriptaddress]{revtex4}

\usepackage[dvips]{graphics}
\usepackage{color}
\definecolor{gold}{rgb}{0.85,0.66,0}
\definecolor{dgreen}{rgb}{0.3,0,0.4}

\begin{document}

\title{\textcolor{dgreen}{Integer Quantum Hall Effect in a Lattice Model
Revisited: Kubo Formalism}}

\author{Paramita Dutta}

\email{paramita.dutta@saha.ac.in}

\affiliation{Theoretical Condensed Matter Physics Division, Saha 
Institute of Nuclear Physics, Sector-I, Block-AF, Bidhannagar, 
Kolkata-700 064, India} 

\author{Santanu K. Maiti}

\email{santanu@post.tau.ac.il}

\affiliation{School of Chemistry, Tel Aviv University, Ramat-Aviv,
Tel Aviv-69978, Israel}

\author{S. N. Karmakar}

\email{sachindranath.karmakar@saha.ac.in}

\affiliation{Theoretical Condensed Matter Physics Division, Saha 
Institute of Nuclear Physics, Sector-I, Block-AF, Bidhannagar, 
Kolkata-700 064, India} 

\begin{abstract}
We investigate numerically the integer quantum Hall effect (IQHE) in a
two-dimensional square lattice with non-interacting electrons in presence 
of disorder and subjected to uniform magnetic field in a direction 
perpendicular to the lattice plane. We employ nearest-neighbor 
tight-binding Hamiltonian to describe the system, and obtain the 
longitudinal and transverse conductivities using Kubo formalism. The 
interplay between the magnetic field and disorder is also discussed. Our 
analysis may be helpful in studying IQHE in any discrete lattice model.
\end{abstract}


\maketitle

\section{Introduction}

The discovery of quantum Hall effect in two-dimensional ($2$D) electron 
systems~\cite{klit,thou} exposed to strong perpendicular magnetic field 
was a triumph of experimental physics. Immediately after this discovery 
scientists from various disciplines were launched into a frenzy of activity 
to understand the underlying physics and also to explore its technological 
importance in designing different 
electronic devices. Such efforts have led to the endowment of a new 
metrological standard, the resistance quantum, $h/e^2$, containing two 
fundamental constants, the electronic charge $e$ and the Planck's constant 
$h$~\cite{hart}. The scaling theory of localization~\cite{tvr} suggests 
that in absence of magnetic field all the states of a $2$D disordered 
non-interacting electron system are localized due to quantum interference. 
The time reversal symmetry is broken in the presence of magnetic field and 
a series of Landau bands appear due to disorder. There are numerous studies
in the literature on the localisation problem of the Landau 
levels~\cite{mac,tan}, specifically studies based on percolation 
theory~\cite{trug} and the calculation of Thouless number~\cite{ando}.
Within each Landau band a central region of extended states is flanked on 
both sides by regions of localized states. For a disordered $2$D electron 
gas it is now well-understood that central region of extended states play 
a significant role towards the quantization of Hall conductance. In the 
limit of strong disorder or weak magnetic field extended states in the 
Landau bands float up in energy~\cite{yang1,yang2}, and, a systematic 
float-up results a transition from integer quantum Hall effect states to 
insulator one~\cite{sheng1}. 

The interplay between magnetic field and disorder is an essential
issue for the phenomenon of IQHE. Breaking of translational invariance 
by impurity potential carries an essential need for the quantization of 
electronic conductivity in presence of magnetic field since in absence 
of disorder no plateau-like structure appears in the Hall conductivity, 
and the classical Hall result is preserved. Appearance of Hall plateaux 
is independent of physical dimension of the sample and this universality 
of the quantum Hall phenomenon helps the technological progress of 
semiconductor physics using 2D electron gas. Not only that, the shape of 
the system is irrelevant in this context. With tremendous experimental 
successes, the phenomenon of IQHE has drawn a lot of interest among 
theoreticians too. Most 
\begin{figure}[ht]
{\centering \resizebox*{4.5cm}{3.5cm}{\includegraphics{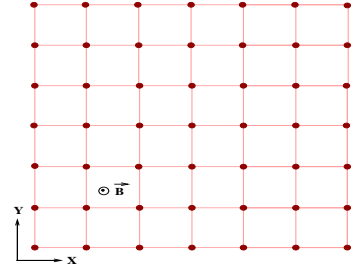}}\par}
\caption{(Color online). Schematic diagram of a square lattice subjected
to a perpendicular magnetic field \mbox{\boldmath$B$}.}
\label{square}
\end{figure}
of the theoretical studies available in the literature are based on the 
continuum model, ignoring the underlying lattice structure of the sample, 
to explain this phenomenon of IQHE. Although there are a few attempts based 
on discrete lattice model~\cite{sheng2,sheng3,czycholl,mandal,san1}, but a 
comprehensive study of this phenomenon with the framework of the lattice 
model is still lacking. In the present article we have made an attempt in 
this direction and study in details the behavior of IQHE in a square lattice 
in presence of disorder and address several important issues.

With this introductory remarks we organize this paper as follows. In section 
II we describe the model and theoretical technique for the calculation of 
longitudinal and transverse conductivities based on Kubo formalism. The 
numerical results and discussion are included in section III. Finally, we 
draw our conclusions in section IV.

\section{Model and Theoretical formulation}

\subsection{The Model}

We start with a $2$D square lattice (Fig.~\ref{square}) which contains 
$N_x$ and $N_y$ number
of atomic sites along the $x$- and $y$-directions, respectively and subject 
it to a perpendicular magnetic field \mbox{\boldmath$B$}. We employ a 
tight-binding (TB) Hamiltonian to describe the system, and, under 
nearest-neighbor hopping approximation it reads as,
\begin{eqnarray}
\mbox{\boldmath{$H$}} &=& \sum_{m,n} \epsilon_{m,n} c_{m,n}^{\dag} c_{m,n} 
+ t \sum_{m,n} \left(c_{m+1,n}^{\dag}c_{m,n} \right. \nonumber \\
& + & \left. c_{m,n+1}^{\dag}c_{m,n}\,e^{i \theta_m} + \mbox{h.c.}
\right)
\label{equ1}
\end{eqnarray}
where, $\epsilon_{m,n}$ denotes the site energy of an electron at site 
($m$,$n$), $m$ and $n$ being the $x$ and $y$ co-ordinates of the site, 
respectively (setting lattice constant $a=1$). In order to incorporate 
impurities in the sample, we choose site energies $\epsilon_{m,n}$ randomly 
from a `Box' distribution function of width $W$. Here, $t$ is the hopping 
integral between the neighboring sites either along $x$- or $y$-direction 
and $c_{m,n}^{\dagger}$ ($c_{m,n}$) is the creation (annihilation) operator 
of an electron at the site ($m$,$n$). In the presence of magnetic field 
\mbox{\boldmath$B$}, a phase factor $\theta_m$ is introduced into the above 
Hamiltonian (Eq.~\ref{equ1}) and for a particular choice of gauge 
\mbox{\boldmath$A$}$\,=$ ($0, Bx$), the so-called Landau gauge, it becomes 
non-zero only when an electron moves along the $y$ direction. Then this phase 
factor can be expressed as $\theta_m=2\pi m \phi$, where $\phi$ is the 
magnetic flux per plaquette measured in units of the elementary flux-quantum 
$\phi_0=ch/e$. We set $\phi=1/Q$, where $Q$ is an integer and the choice of 
$Q$ should be such that it is commensurate with $N_y$. This reduces the 
boundary conditions to the traditional periodic ones~\cite{sheng4}. 

\subsection{Linear Response Kubo Formalism}

To obtain the longitudinal and transverse conductivities, we use the Kubo 
formalism which is briefly outlined below. The general expression for
electrical conductivity is written in the form,
\begin{widetext}
\begin{eqnarray}
\sigma_{kl}& =& \frac{i e^2 \hbar}{N} \sum_{\alpha}\sum_{\beta \neq \alpha}
\left(f_{\alpha}-f_{\beta}\right) 
\frac{\langle \alpha|\dot{\mbox{\boldmath{$u$}}}_k|\beta \rangle \langle 
\beta |\dot{\mbox{\boldmath{$u$}}}_l| \alpha \rangle}
{\left(\mathcal{E}_{\alpha}-\mathcal{E}_{\beta}\right)^2 + \eta^2} 
+ \frac{e^2 \hbar}{N} \sum_{\alpha} \sum_{\beta \neq \alpha} 
\left(\frac{f_{\alpha}-f_{\beta}}{\mathcal{E}_{\alpha}-\mathcal{E}_{\beta}}
\right) 
\frac{\eta}{\left(\mathcal{E}_{\alpha}-\mathcal{E}_{\beta}\right)^2 + 
\eta^2} \langle \alpha|\dot{\mbox{\boldmath{$u$}}}_k|\beta \rangle
\langle \beta |\dot{\mbox{\boldmath{$u$}}}_l| \alpha \rangle 
\label{sigij}
\end{eqnarray}
\end{widetext}
where, $\eta \rightarrow 0^+$. Here, the indices $k$ and $l$ can be $x$ or 
$y$. For $k=l=x$, we get $\sigma_{xx}$, the so-called longitudinal 
conductivity while for the other case we have the transverse conductivity, 
$\sigma_{xy}$. The states $|\alpha \rangle$ and $|\beta \rangle$ are the 
eigenstates of the Hamiltonian (Eq.~\ref{equ1}) corresponding to the energy 
eigenvalues $\mathcal{E}_{\alpha}$ and $\mathcal{E}_{\beta}$, respectively, 
and $N=N_x\times N_y$ represents the size of the sample. 
$\dot{\mbox{\boldmath{$u$}}}_k$ is the velocity operator along $k$-th ($x$ 
or $y$) direction and $f_{\alpha(\beta)}=
1/[1+Exp\left\{(\mathcal{E}_{\alpha(\beta)}-E_F)/k_B T)\right\}]$ is
the Fermi distribution function at absolute temperature $T$ with Fermi energy 
$E_F$.

When $k$ and $l$ are identical to each other, the factor
$\langle \alpha|\dot{\mbox{\boldmath{$u$}}}_k|\beta \rangle \langle 
\beta |\dot{\mbox{\boldmath{$u$}}}_l| \alpha \rangle$ in Eq.~\ref{sigij}
becomes a real and positive one, while it becomes purely imaginary when
$k$ and $l$ are different. Since the conductivity is a real and positive
quantity, the first term on the right hand side of Eq.~\ref{sigij} gives
transverse conductivity ($\sigma_{xy}$), while the second term represents
longitudinal conductivity ($\sigma_{xx}$). Separating the transverse and 
longitudinal parts we can write,
\begin{equation}
\sigma_{xy}=\frac{i e^2 \hbar}{N} \sum_{\alpha}\sum_{\beta \neq \alpha}
(f_{\alpha}-f_{\beta}) ~\frac{\langle \alpha|\dot{\mbox{\boldmath{$u$}}}_x|
\beta \rangle \langle \beta |\dot{\mbox{\boldmath{$u$}}}_y| \alpha \rangle}
{(\mathcal{E}_{\alpha}-\mathcal{E}_{\beta})^2 + \eta^2}
\label{sigxy}
\end{equation}
and 
\begin{equation}
\sigma_{xx}=\frac{e^2 \hbar}{N} \sum_{\alpha} \sum_{\beta \neq \alpha} 
\left(\frac{f_{\alpha}-f_{\beta}}{\mathcal{E}_{\alpha}-\mathcal{E}_{\beta}}
\right) \frac{\eta}{\left(\mathcal{E}_{\alpha}-\mathcal{E}_{\beta}\right)^2 
+ \eta^2}|\langle \alpha|\dot{\mbox{\boldmath{$u$}}}_x|\beta \rangle |^2.   
\label{sigxx}
\end{equation}
\vskip 0.3cm
\noindent
$\bullet$ {\underline{\bf Matrix elements of the Velocity operators:}}
\vskip 0.3cm
\noindent
To calculate the matrix elements of the velocity operators
$\dot{\mbox{\boldmath{$u$}}}_k$ we start with the basic relation~\cite{our},
\begin{equation}
\dot{\mbox{\boldmath{$u$}}}_k=\frac{1}{i \hbar}\left[\mbox{\boldmath{$u_k$}},
\mbox{\boldmath{$H$}}\right]
\label{velopr}
\end{equation}
where, $\mbox{\boldmath{$u$}}_k$ is the displacement operator along $k$-th 
($x$ or $y$) direction and \mbox{\boldmath{$H$}} is the Hamiltonian operator
described in Eq.~\ref{equ1}. Expanding and simplifying the above relation
(Eq.~\ref{velopr}) we get the velocity operators in the following second 
quantized form along the $x$- and $y$-directions, 
\begin{equation}
\dot{\mbox{\boldmath{$u$}}}_x=\frac{i t}{\hbar}\sum_{m,n}\left(c_{m,n}^{\dag} 
c_{m+1,n}-c_{m+1,n}^{\dag}c_{m,n}\right)
\label{vx}
\end{equation}
and,
\begin{equation}
\dot{\mbox{\boldmath{$u$}}}_y=\frac{i t}{\hbar}\sum_{m,n}\left(c_{m,n}^{\dag} 
c_{m,n+1}\, e^{-i \theta_m} - c_{m,n+1}^{\dag}c_{m,n}\, e^{i \theta_m}\right).
\label{vy}
\end{equation}
Therefore, the {textcolor{red}{matrix elements of the velocity operators} with respect to the 
eigenvectors $|\alpha \rangle$ and $|\beta \rangle$ become,
\begin{equation}
\langle \alpha |\dot{\mbox{\boldmath{$u$}}}_x| \beta \rangle = 
\frac{i t}{\hbar} \sum_{m,n}\left(a^{\alpha~*}_{m,n}\, a^{\beta}_{m+1,n}-
a^{\alpha~*}_{m+1,n}\, a^{\beta}_{m,n}\right)
\label{equ10}
\end{equation}
and,
\begin{eqnarray}
\langle \alpha |\dot{\mbox{\boldmath{$u$}}}_y| \beta \rangle & =& 
\frac{i t}{\hbar} \sum_{m,n} \left(a^{\alpha~*}_{m,n}\, a^{\beta}_{m,n+1} 
\,e^{-i \theta_m}\right. \nonumber \\ 
& - & \left.a^{\alpha~*}_{m,n+1}\, a^{\beta}_{m,n}
\, e^{i \theta_m}\right)
\label{equ11}
\end{eqnarray}
where, the eigenvectors look like,
\begin{equation}
|\alpha \rangle =\sum_{p,q} a^{\alpha}_{p,q} \, |p,q \rangle
\end{equation}
and,
\begin{equation}
|\beta \rangle =\sum_{p,q} a^{\beta}_{p,q} \, |p,q \rangle.
\end{equation}
Here, $|p,q\rangle$'s are the Wannier states and $a^{\alpha}_{p,q}$ and 
$a^{\beta}_{p,q}$'s are the corresponding coefficients. Taking the 
contributions from all these states we find the matrix elements of the 
velocity operators along the $x$- and $y$-directions. 

Substituting the above expressions (Eqs.~\ref{equ10} and \ref{equ11}) 
into Eqs.~\ref{sigxy} and~\ref{sigxx} we get the relations,
\begin{eqnarray}
\sigma_{xy} & = & -\frac{2 \pi i t^2}{hN} \sum_{\alpha} 
\sum_{\beta \neq \alpha} \frac{\left(f_{\alpha}-f_{\beta}\right)}
{\left(\mathcal{E}_{\alpha}-\mathcal{E}_{\beta}\right)^2+\eta^2} 
\nonumber \\
& \times & \left\{\sum_{m,n} \left(a^{\alpha~*}_{m,n} a^{\beta}_{m+1,n} - 
a^{\alpha~*}_{m+1,n} a^{\beta}_{m,n}\right)\right\} \nonumber \\
& \times & \left\{\sum_{m,n} \left(a^{\beta~*}_{m,n} a^{\alpha}_{m,n+1}
e^{-i \theta_m} - a^{\beta~*}_{m,n+1} a^{\alpha}_{m,n} 
e^{i \theta_m}\right)\right\} \nonumber \\
\label{sigmodxy}
\end{eqnarray}
and,
\begin{eqnarray}
\sigma_{xx} & = & \frac{2 \pi t^2}{hN} \sum_{\alpha} 
\sum_{\beta \neq \alpha} \left(\frac{f_{\alpha}-f_{\beta}}
{\mathcal{E}_{\alpha}-\mathcal{E}_{\beta}}\right)
\frac{\eta}{\left(\mathcal{E}_{\alpha}-\mathcal{E}_{\beta}\right)^2+\eta^2}
\nonumber \\
& \times & \left|\sum_{m,n}\left(a^{\alpha~*}_{m,n} a^{\beta}_{m+1,n}- 
a^{\alpha~*}_{m+1,n} a^{\beta}_{m,n}\right)\right|^2.
\label{sigmodxx}
\end{eqnarray}
These are the final expressions for the transverse and longitudinal 
conductivities, respectively, which offer a very convenient method for 
numerical calculation of the conductivities using Kubo formalism.

Throughout our study we choose the units where $c=e=h=1$ and measure the
energy scale in units of $t$.

\section{Numerical results and discussion}

\subsection{Energy Spectrum}

To make this present communication a self-contained one let us begin
with the energy band structure of our model quantum system. 

In Fig.~\ref{level} we present the energy levels for a finite size square 
lattice in presence of a transverse magnetic field both for the ordered as 
well as the disordered cases. The vertical lines correspond to the locations 
of the energy eigenvalues. For the ordered case we use black lines, while 
the blue lines are used for the disordered case, and, these two colored lines 
are superimposed in each plot to reveal the effect of disorder clearly.  
In the absence of disorder we get very sharp lines associated with the
energy eigenvalues of the lattice and all these energy levels are highly
degenerate. The degeneracy factor of the energy levels strongly depends
on the specific choice of the magnetic field, i.e., on the value of $Q$. 
When $Q$ becomes commensurate with $N_y$, we get $Q$ number of such sharp 
\begin{figure}[ht]
{\centering \resizebox*{6.5cm}{4.3cm}
{\includegraphics{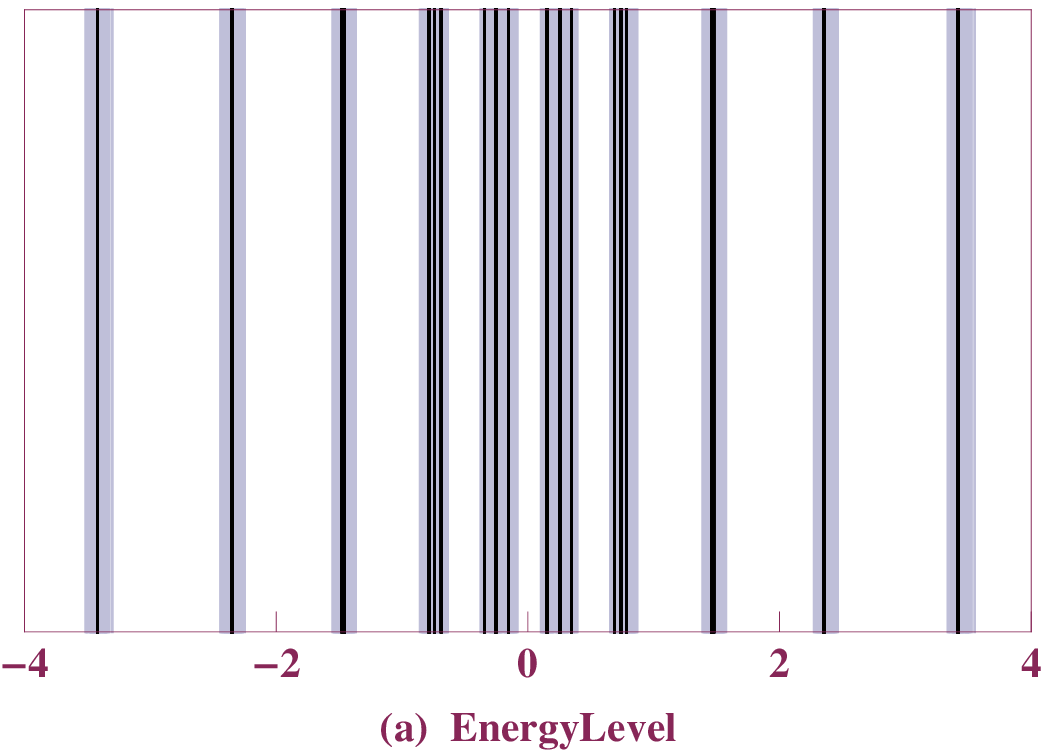}}\par}
{\centering \resizebox*{6.5cm}{4.3cm}
{\includegraphics{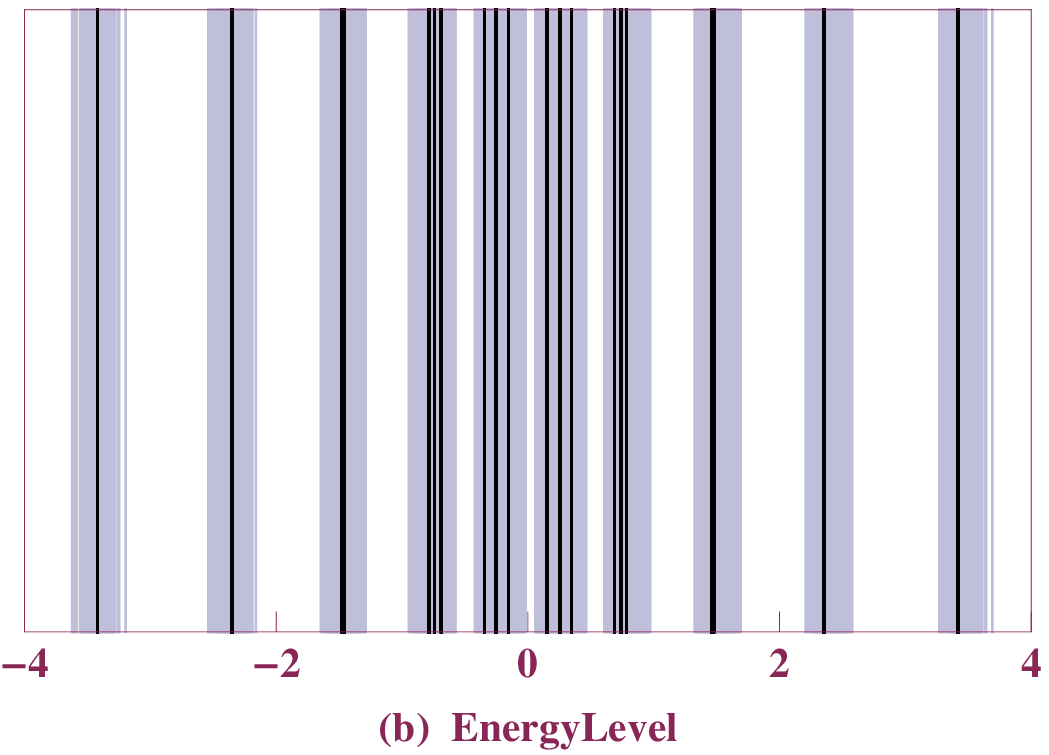}}\par}
{\centering \resizebox*{6.5cm}{4.3cm}
{\includegraphics{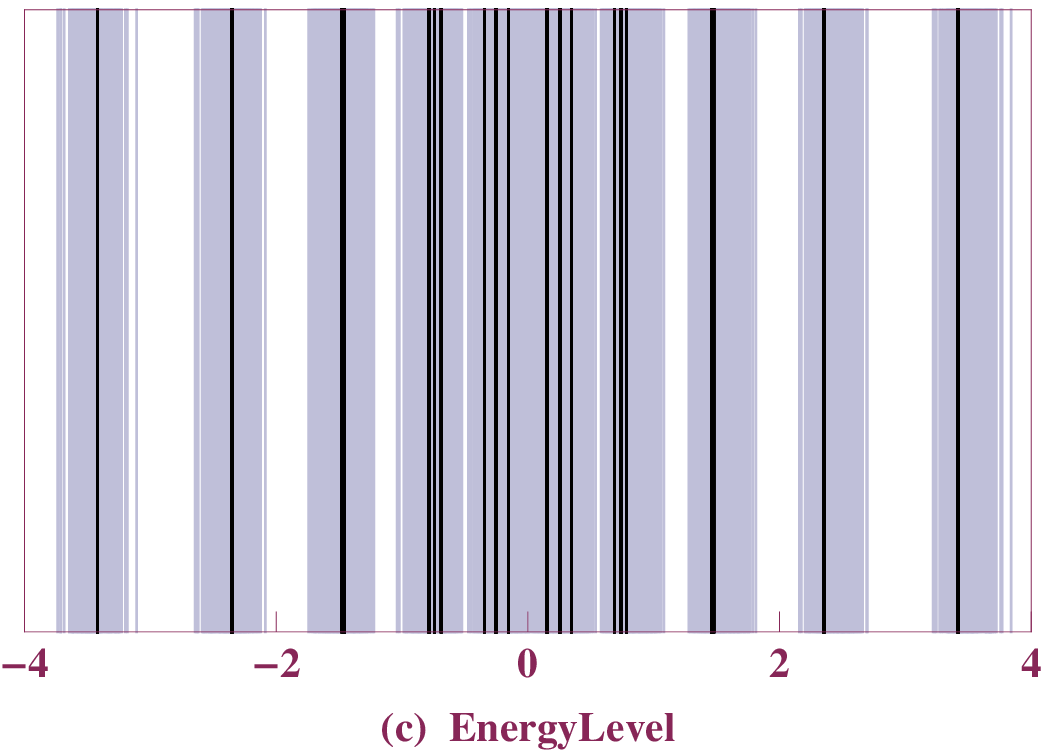}}\par}
\caption{(Color online). Energy levels for a square lattice ($30 \times 30$) 
when $\phi$ is set at $0.1$. The black lines represent the locations of the
energy levels for the ordered ($W=0$) case. On the other hand, the blue lines 
denote the positions of the energy levels for the disordered ($W \ne 0$) case,
where (a), (b) and (c) correspond to $W=1$, $2$ and $3$, respectively.}
\label{level}
\end{figure}
energy levels and each level becomes $N/Q$-fold degenerate. These levels 
are quite similar to the Landau levels of the continuum model of 2D electron 
gas. It is to be noted that unlike the equispaced Landau levels of the 
continuum model, we do not have equispaced energy levels for finite sized 
squarelattice. The levels are widely separated near the two edges of the 
energy spectrum as observed from Fig.~\ref{level}. The situation is somewhat 
interesting when impurities are introduced into the system. 
In presence of 
disorder the degeneracy of each Landau band gets lifted due to the impurity 
potentials and the sharp energy levels form quasi-bands keeping the total 
number of energy levels in each quasi-band identical to each other and this 
number is equal to the degeneracy of the band. As far example, if we choose 
$N_x=N_y=30$ and set the magnetic flux $\phi$ at $0.1$ (i.e., $Q=10$), then 
$10$ Landau bands appear in the spectrum (Fig.~\ref{level}) and each band 
contains $90$ energy levels. The width of each Landau band increases with 
the increase of strength of disorder $W$ as can be observed from the spectra, 
and for large enough disorder strength the neighboring Landau bands start to 
overlap with each other.

\subsection{Transverse and Longitudinal Conductivities}

Let us now illustrate the characteristics features of longitudinal 
and transverse conductivities and the related issues for a finite
sized square lattice. In Fig.~\ref{sigmaxy} we show the variation of 
transverse conductivity $\sigma_{xy}$ (red line) as a function of 
Fermi energy $E_F$ for a ($60 \times 60$) square lattice considering
$\phi=0.1$, $W=1$ and $k_BT=0.01$
\begin{figure}[ht]
{\centering \resizebox*{7.5cm}{5cm}{\includegraphics{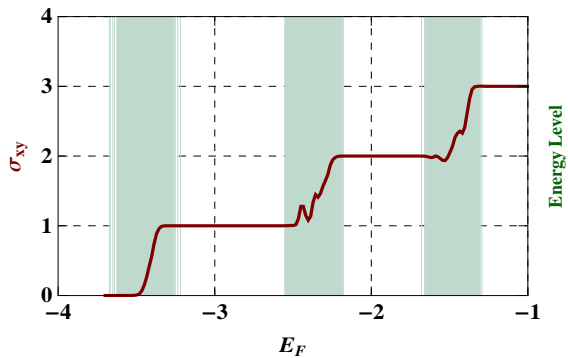}}\par}
\caption{(Color online). Transverse conductivity (red curve) as a function 
of Fermi energy for a square lattice ($60 \times 60$) considering $\phi=0.1$, 
$W=1$ and $k_BT=0.01$. The Landau bands (light green) are superimposed on it.}
\label{sigmaxy}
\end{figure} 
The Landau levels (light green) are also superimposed on the spectrum. 
From the spectrum it is clear that the transverse/Hall conductivity 
$\sigma_{xy}$ increases in integer steps with the rise of Fermi energy
$E_F$ and at these plateaux the Hall conductivity gets the value, $\nu e^2/h$
with a great accuracy, where the integer number $\nu$ corresponds to the 
total number of filled Landau bands below the Fermi level. For a particular 
system size the total number of available Hall plateaux strongly depends on
the choice of magnetic flux $\phi$, since the specific choice of $\phi$ fixes 
the total number of Landau bands as mentioned earlier. Very interestingly
we observe that even for a square lattice whenever the Fermi energy crosses 
any one of the Landau bands, the Hall conductivity enhances exactly by the 
same amount $e^2/h$, and accordingly, for $\nu$ filled Landau bands, 
$\sigma_{xy}$ becomes equal to $\nu e^2/h$, which is precisely the 
characteristic feature of integer quantum Hall effect. It is important to 
note that in traditional ballistic
waveguides we also get quantized conductance, but such a strange precise
quantization, about one part in million, can never be acheived in ordinary 
ballistic conductors as the backscattering processes are not completely 
suppressed. The almost complete elimination of the backscattering processes 
can only be obtained in the quantum Hall regime which actually ensures the 
extremely high precision in the quantized value of Hall conductivity. Recently,
it has been reported that when the Fermi energy lies within a plateau region,
there is practically zero overlap between the current carrying states in the 
\begin{figure}[ht]
{\centering \resizebox*{7.5cm}{5cm}{\includegraphics{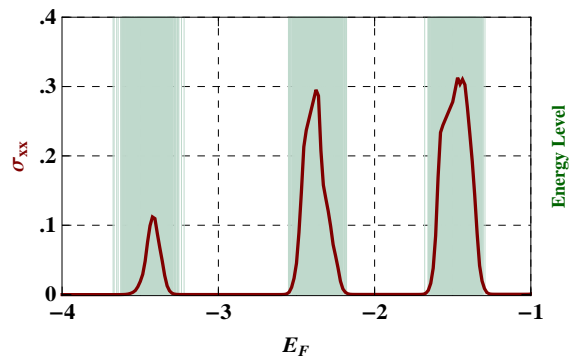}}\par}
\caption{(Color online). Longitudinal conductivity (red curve) as a function 
of Fermi energy for a square lattice ($60 \times 60$) for the same parameter
values given in Fig.~\ref{sigmaxy}. The Landau bands (light green) are 
superimposed on it.}
\label{sigmaxx}
\end{figure}
sample leading to the almost complete suppression of the backscattering 
processes and thereby the momentum relaxation process~\cite{san1}. 

As a result of almost entire suppression of the backscattering processes
for Fermi energy lying in the plateau regions, we would get practically 
zero longitudinal resistance and hence vanishing longitudinal conductivity
\begin{figure*}[ht]
{\centering \resizebox*{5.5cm}{10.5cm}{\includegraphics{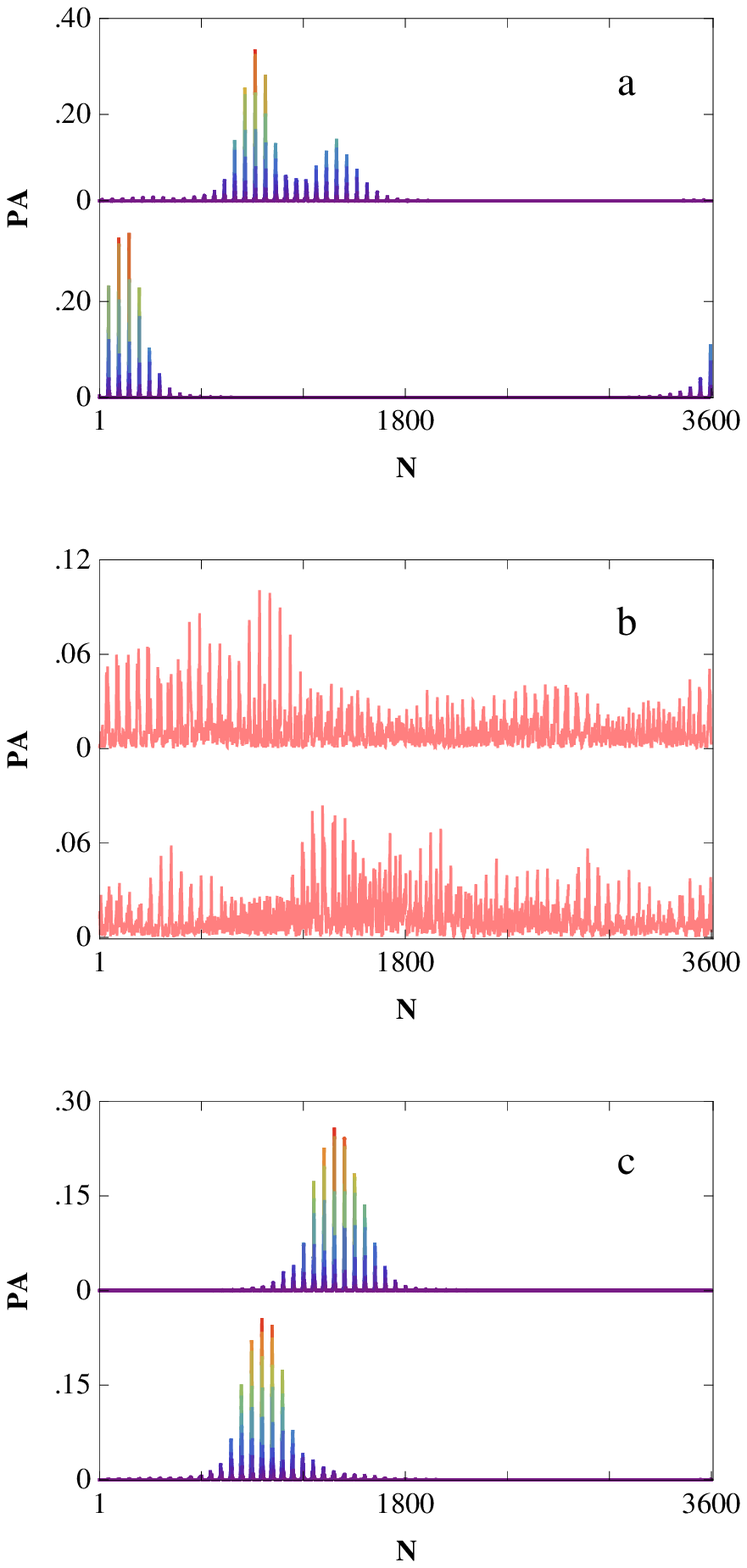}}
\resizebox*{5.5cm}{10.5cm}{\includegraphics{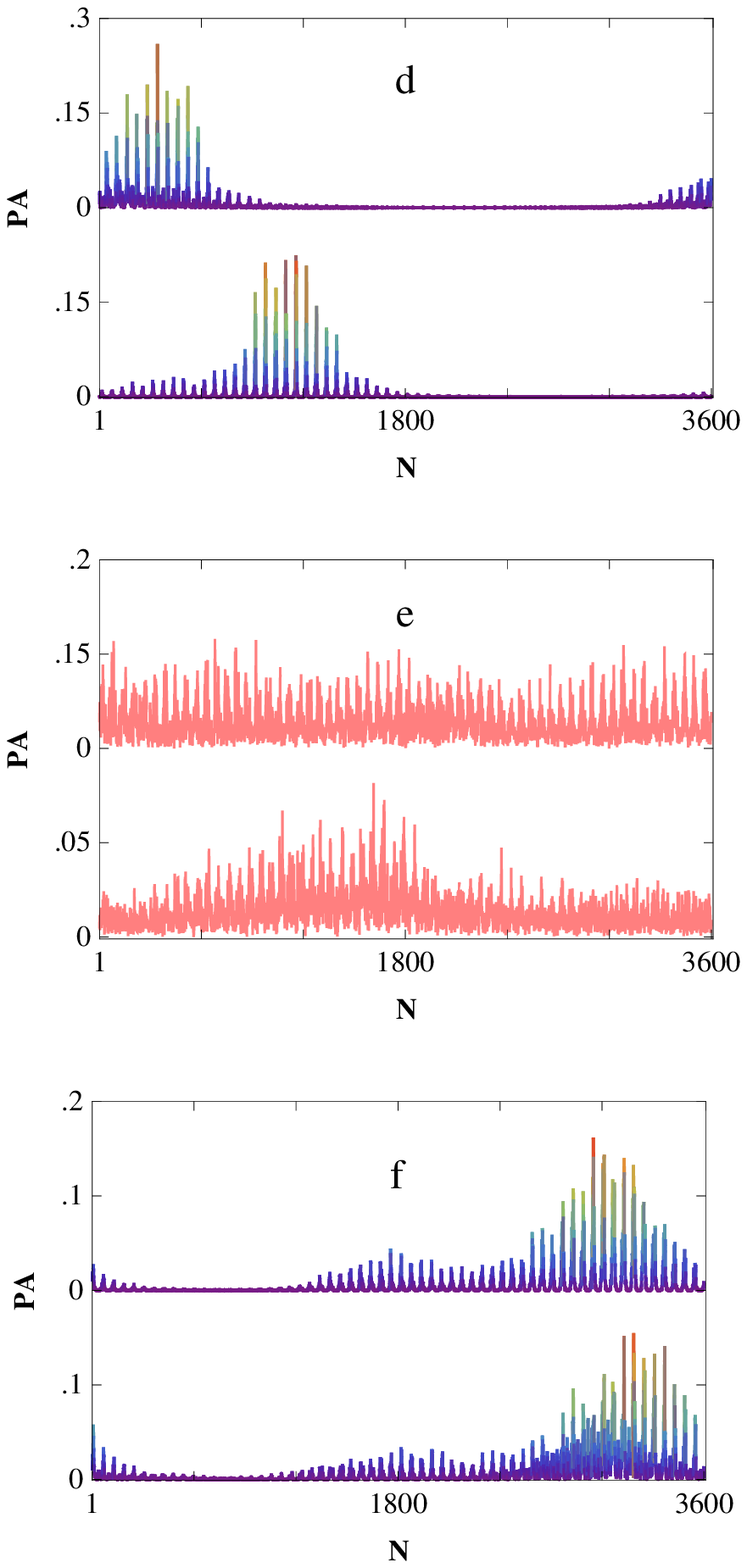}}
\resizebox*{5.5cm}{10.5cm}{\includegraphics{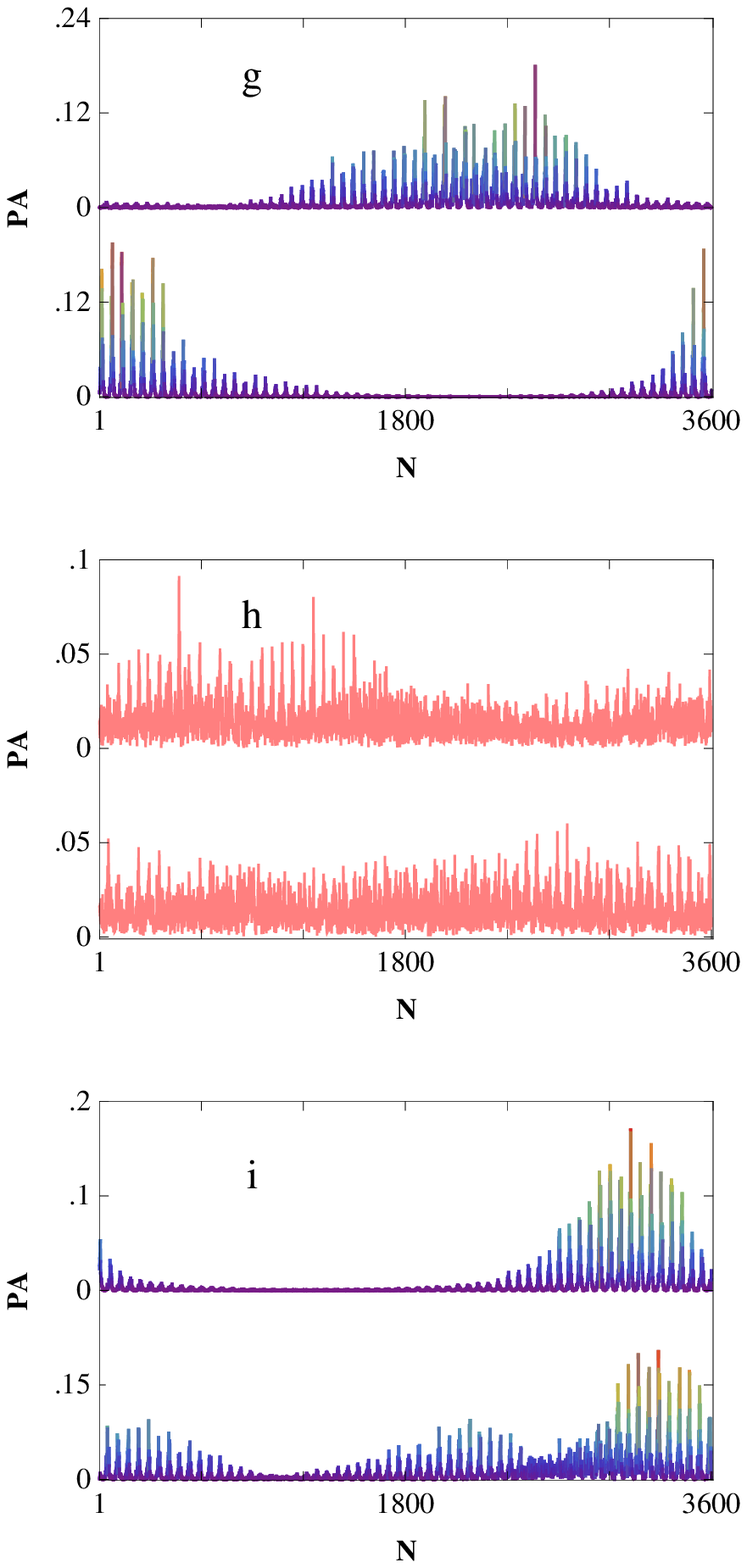}}\par}
\caption{(Color online). Probability amplitude (PA) at different lattice
sites (N) of a square lattice ($60 \times 60$) for the same parameter 
values as mentioned in Fig.~\ref{sigmaxy}. The top and bottom panels 
correspond to the results for the eigenstates selected from the left 
and right edges of the Landau band, respectively, while the middle one 
is for the eigenstates lie in the centre of the Landau band. The three 
different columns are associated with the three different Landau bands 
shown in Fig.~\ref{sigmaxy}. To understand the nature of energy eigenstates
more clearly, in each figure we present the results for two states those
are chosen from the respective regions.}
\label{PA}
\end{figure*}
which directly follows from the conductivity tensor. The result is shown
in Fig.~\ref{sigmaxx} where we set the same parameter values as given in
Fig.~\ref{sigmaxy}. The resistance of a conductor is associated with the 
rate at which electrons can relax their momentum. To loose the momentum 
an electron has to be scattered through the allowed energy eigenstates in
the bulk of the conductor. This almost vanishes whenever the Fermi energy
lies between two Landau bands. The longitudinal resistance is non-vanishing
only when the Fermi energy is within a Landau band where the backscattering 
process are present.  

Figure~\ref{sigmaxy} 
clearly shows that the Hall plateaux get extended well
inside the Landau bands and the Hall conductivity rises only within certain
central region of each Landau bands. The underlying physics behind it as 
follows. In presence of disorder Landau bands get broadened and the natures 
of all eigenstates in a Landau band are not identical. The states those lie 
along the two edges of the Landau band are localized, while the states in 
the middle are almost extended. To reveal the nature of the energy 
eigenstates whether they are quasi-localized or extended, in Fig.~\ref{PA} 
we show the variation of probability amplitude at different lattice sites of 
a square lattice with the same set of parameters as those taken in 
Fig.~\ref{sigmaxy}. The top and bottom panels correspond to the energy levels 
chosen from the left and right edges of the three Landau bands, respectively, 
while the middle panel refers to some of the energy levels near the centre of 
the Landau bands. Quite clearly we see that near the band edges the 
probability amplitude vanishes almost at all sites excepting a few, which 
implies that electrons are localized in these states. On the other hand, for 
the states well inside the Landau bands we have finite probability amplitude 
on each site revealing the extended nature of the states. This behavior is 
true for every Landau band. Therefore, for the energy levels which lie near 
the two edges of the Landau bands we have almost zero contribution to the 
transverse conductivity and it leads to the plateau regions in the Hall 
conductivity, and, the non-zero contribution comes only from the extended 
states near the center of the Landau bands. We also observe that the length 
of the Hall plateaux get widened with the rise of impurity strength, but it 
does not affect the quantized nature of Hall conductance. The quantization of 
Hall conductance is very robust and really very surprising. This is due to 
the fact that when we increase the strength of disorder in the sample, more 
states become localized from the edges of the Landau bands and the regions of 
extended states in the band centre becomes narrow. However, the contribution 
to Hall conductivity that gets lost due to these additional localized states 
is exactly compensated by the enhanced contribution from the extended states, 
and accordingly, the quantized nature of Hall conductivity is remains 
unaltered. This phenomenon is known as the current compensation and it has 
already been discussed in the literature by using the continuum 
model~\cite{janssen}.

\subsection{Effect of Magnetic Field and Temperature}

So far we have discussed the main aspects of integer quantum Hall effect 
in a square lattice, but for the sake of completeness now we focus our 
attention on the effects of magnetic field and temperature on IQHE as this 
phenomenon is highly sensitive to the interplay of these parameters. With 
the increase of disorder strength the Hall plateaux can be found to be 
destroyed one by one in a generic sequence~\cite{sheng2}.

In Fig.~\ref{fld} we present the variation of Hall conductivity of a square 
lattice ($60 \times 60$) as a function of Fermi energy 
\begin{figure}[ht]
{\centering \resizebox*{7.25cm}{4.75cm}{\includegraphics{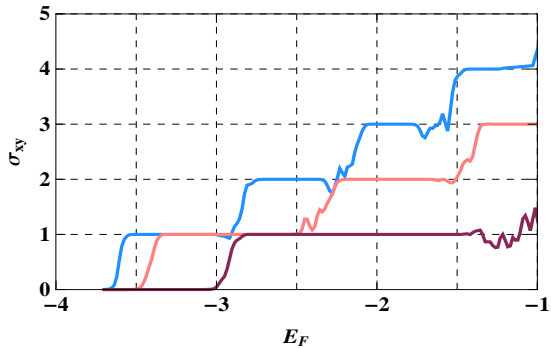}}\par}
\caption{(Color online). Hall conductivity as a function of Fermi energy 
for a square lattice ($60 \times 60$) with $W=1$ and $K_BT=0.1$. 
The dark-violet, pink and blue lines correspond to $Q=5$, $10$ and $15$, 
respectively.}
\label{fld}
\end{figure}
\begin{figure}[ht]
{\centering \resizebox*{7.25cm}{4.75cm}{\includegraphics{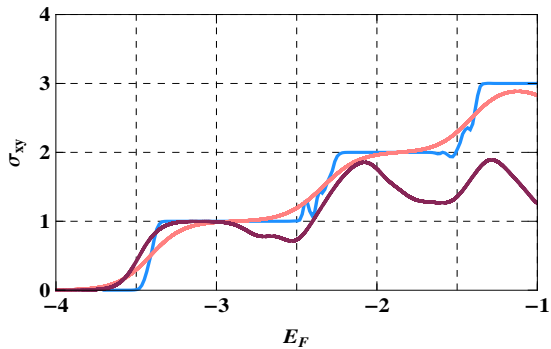}}\par}
\caption{(Color online). Hall conductivity as a function of Fermi energy 
for a square lattice ($60 \times 60$) considering $W=1$ and $\phi=0.1$. 
The blue, pink and dark-violet colors correspond to $k_BT=0.01$, $0.05$, 
and $0.1$, respectively.}
\label{tem}
\end{figure}
for three different values of magnetic flux $\phi$ regulated by the parameter 
$Q$ since $\phi=1/Q$. Here we
set $k_BT=0.1$ and $W=1$. The dark-violet, pink and blue curves correspond
to $Q=5$, $10$ and $15$, respectively, and these values of $Q$ are such
that $N_x$ is exactly divisible by them so that the translational invariance
remains along the $x$-direction. For a given value of $Q$, we have $Q$ Landau 
bands in the energy spectrum, and accordingly, for a certain energy range 
the number of Hall plateaux are different for various $Q$ as can be seen 
clearly from Fig.~\ref{fld}. However, in all these cases the quantized nature 
of the Hall conductance is exactly maintained, i.e., $\sigma_{xy}$ is integer 
multiple of the factor $e^2/h$ with a great accuracy.

The effect of temperature in IQHE is illustrated in Fig.~\ref{tem}, where we 
show the results for three temperatures considering $W=1$ and flux $\phi=0.1$. 
For a finite disorder strength and weak magnetic field, the widths of the Hall 
plateaux decrease gradually with the rise of temperature and the conductance 
spectrum becomes quite messy. For a sufficiently large temperature the
phenomenon of conductance quantization almost disappears and this phenomenon 
has also been verified experimentally by some groups~\cite{cage,wei}. When 
the Fermi level is placed anywhere within the localized regions, the thermally 
excited electrons can jump toward the extended region and contribute to the 
conductivity. Thus, there is a competition between these two issues as a resultthe quatization gets lost gradually.

\section{Conclusion}

In conclusion, we have re-visited the phenomenon of integer quantum Hall
effect in a two-dimensional square lattice within a non-interacting electron
picture. The interplay between the magnetic filed and randomness has been
discussed in detail. We have used a single-band nearest-neighbor tight-binding 
Hamiltonian to describe the model quantum system and numerically calculated 
the longitudinal and transverse conductivities by using Kubo formalism. The
physical picture about the integer quantum Hall effect that has emerged 
from our present investigation based on the discrete lattice model is almost
the same as that obtained in the continuum model. Our approach could be much 
more suitable for further investigation of the Quantum Hall effect in 
topologically insulating materials, like, graphene flakes, kagome lattices, 
etc. This is our first step towards this direction.

\section{Acknowledgment}

We thank Shreekantha Sil for some useful discussion.

\end{document}